\pdfoutput=1
\documentclass[10pt,a4paper]{article}

\usepackage{amsmath}
\usepackage{amssymb}
\usepackage{amsthm}
\usepackage{layouts}
\usepackage{cite}
\usepackage{hyperref}

\usepackage{microtype}
\DisableLigatures[f]{encoding = *, family = * }
\usepackage{graphicx}

\usepackage[labelfont=bf,labelsep=period,justification=raggedright]{caption}

\makeatletter
\renewcommand{\@biblabel}[1]{\quad#1.}
\makeatother

\date{}

\newcommand{\ket}[1]{\ensuremath{|#1\rangle}}
\newcommand{\bra}[1]{\ensuremath{\langle#1|}}
\newcommand{\ketbra}[2]{\ensuremath{\ket{#1}\bra{#2}}}

\newcommand{\HH}{\mathcal{H}}
\newcommand{\Tr}{\mathrm{Tr}}
\newcommand{\1}{{\rm 1\hspace{-0.9mm}l}}
\newcommand{\ii}{\mathrm{i}}
\newcommand{\LL}{\mathcal{L}}

\newcommand{\TOM}[1]{\mathcal{#1}}
\newcommand{\ie}{{\emph{i.e.\/}}}

\newtheorem{theorem}{Theorem}
\newtheorem{definition}{Definition}
\newtheorem{remark}{Remark}
\newtheorem{proposition}{Proposition}

\newenvironment{theproof}[1][Proof]{\noindent\textbf{#1.} 
}{\hfill\rule{0.5em}{0.5em}}

\begin{document}

\begin{flushleft}
{\Large
\textbf{Generalized open quantum walks on Apollonian networks}
}
\\
{\L}. Pawela$^{1,\ast}$, 
P. Gawron$^{1}$, 
J.A. Miszczak$^{1}$
P. Sadowski$^{1}$
\\
\bf{1} Institute of Theoretical and Applied Informatics,
Polish Academy of Sciences,
Ba{\l}tycka 5, 44-100 Gliwice, Poland
\\
$\ast$ E-mail: Corresponding lpawela@iitis.pl
\end{flushleft}

\section*{Abstract}
We introduce the model of generalized open quantum walks on networks using the
Transition Operation Matrices formalism. We focus our analysis on the mean
first passage time and the average return time in Apollonian networks. These
results differ significantly from a classical walk on these networks. We show a
comparison of the classical and quantum behaviour of walks on these networks.
\section*{Introduction}\label{sec:introduction}
Understanding the information flow in classical and quantum networks is crucial
for the comprehension of many phenomena in physics, social sciences and
biology~\cite{barabasi99emergence, watts99small, albert02statistical}.
Real-world networks are usually small-world and scale-free. An important example
of networks which posses both of these properties are Apollonian networks.
Random walks provide a useful model for studying the behaviour of agents in
complex networks~\cite{jansen02intrusion, bernardes00implementation,
lange99seven, miszczak2014magnus,miszczak2014general, pawela2013cooperative,
pawela2013quantum}.

In particular in \cite{noh2004random} it was shown that for the class of finite
connected undirected networks, walks for which probability of leaving a node is
reciprocal of its degree, have a fixed average return time (ART). Mean first
passage time (MFPT) and ART in the case of Apollonian networks have been studied
in \cite{huang05walks}.

In this paper we investigate the behaviour of quantum walks of the class of
Apollonian networks. Using the concept of generalised open quantum walks (GOQW)
we introduce the definition of MFPT in the quantum case. The notion of GOQW
allows us the consideration of a broader class of walks compared to the open
quantum walks introduced in \cite{attal12open, attal2012open, sinayskiy2013open,
sweke2013dissipative, sinayskiy2012properties, sinayskiy2014quantum,
attal2012central, sadowski2014central}. The main limitation in using the open
quantum walks is the lack of flexibility in assigning the weights to the edges.

The motivation for performing the research presented in this paper was to study
coin-less quantum walks on undirected graphs, with weights on edges. We also
assume that for each edge its weight in one direction is not necessary related
with the weight in the other direction. In the usual setting
\cite{kempe2003quantum} quantum walks are defined by a Hamiltonian derived from
the adjacency matrix. Due to the hermiticity of the Hamiltonian the intensity of
transition from vertex $i$ to vertex $j$ is related to the intensity of
transition from $j$ to $i$. One way to overcome this limitation is to use the
technique introduced by Szegedy \cite{szegedy2004quantum}. In this paper we
propose an alternative approach by introducing generalized open quantum walks.
Moreover, we apply this formalism to extend the analysis performed in
\cite{noh2004random}, where the authors studied the relation between the degree
of the vertex and mean first passage time of a Markov process.

The paper is organized as follows. In the next section we introduce basic
concepts concerning the presented work such as the notions of quantum mechanics,
a generalization of the open quantum walk (OQW) model and the notion of quantum
transition operation matrix (TOM). Subsequently we provide the methods of
constructing the generalized open quantum walks on Apollonian networks and
discuss some particular cases. Finally, we provide the concluding remarks and
suggest a direction for further work.

\section*{Preliminaries}\label{sec:prelim}
\subsection*{Apollonian networks}\label{sec:apollonian}
Apollonian networks are named after Apollonius of Perga, who introduced the
problem of space filling by packing
spheres~\cite{boyd73osculatory,boyd73residual}. The concept of Apollonian
networks was introduced by Andrade \emph{et al.} \cite{andrade05apollonian} and
by Doye and Massen in \cite{doye05self-similar}. In \cite{andrade05apollonian}
it was shown that it can be used to describe force chains in polydisperse
granular packings, whilst in \cite{doye05self-similar} topological and spatial
properties of such networks are characterized and their application as model for
networks of connected energy minima is discussed.

Apollonian networks display some properties which make them a very useful tool
for studying effects in large complex networks. In particular they have the
property of being scale-free and small-world. They can be also embedded in
Euclidean lattice, and show space filling and matching graph properties.

Apollonian networks have been used in various areas of science. In particular,
Andrade and Herrmann \cite{andrade05magnetic} and Serva \emph{et al.}
\cite{serva13ising} investigated the properties of Ising models on Apollonian
network. It was also suggested that Apollonian networks can be harnessed to
mimic a behaviour of neuronal systems in the brain~\cite{pellegrini07modelling}.
Random Apollonian networks \cite{zhou04random} were introduced as a model for
real-world planar graphs. Their high-dimensional generalizations were also
proposed in~\cite{zhang06high-dimensional}. The properties of random Apollonian
networks were studied in \cite{frieze12certain} in the context of web graphs.

\begin{figure}[!h]
\centering
\includegraphics[width=0.6\textwidth]{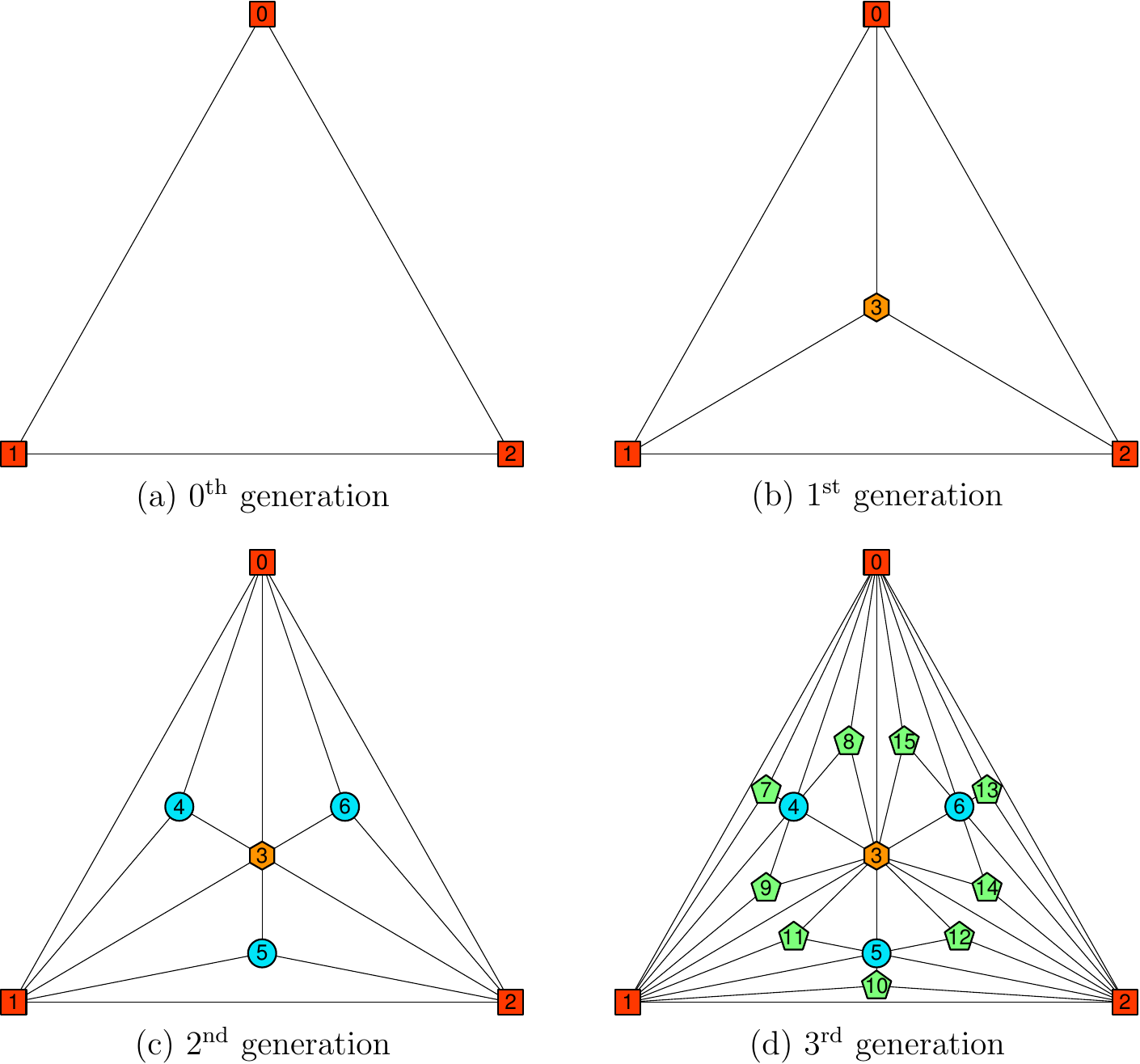}
\caption{ {\bf An illustration of the construction of an Apollonian network.}
Red squares illustrate the nodes in the 0\textsuperscript{th} generation, an
orange hexagon in the 1\textsuperscript{st} generation, blue circles in the
2\textsuperscript{nd} generation and green pentagons in the
3\textsuperscript{rd} generation.}
\label{fig:example-network}
\end{figure}

The construction of a regular Apollonian network can be done by a~recursive
procedure. At first, a complete 3-vertex graph is created, we call it the
0\textsuperscript{th} generation Apollonian network. In order to obtain the next
generation network new nodes are inserted in the middle of each of the triangles
in the graph. Each of the new vertices is associated with three new edges
connecting it to the vertices of the corresponding triangle. Apollonian networks
of generations zero to three resulting from the algorithm are presented in the
Fig.~\ref{fig:example-network}.

Various researchers considered walks on Apollonian networks. For example Huang
\emph{et al.} \cite{huang05walks} studied classical random walks on
deterministic and random Apollonian networks. Random walks on Apollonian
networks with defects were considered by Zang \emph{et al.}
\cite{zhang09randomwalks}. Discrete time quantum walks on Apollonian networks
were studied by Souza and Andrade in \cite{souza13discrete}, where a comparison
of the introduced model with its classical counterpart was provided. Xu \emph{et
al.} \cite{xu08coherent} studied the properties of coherent exciton transport on
Apollonian networks with dynamics modelled by continuous-time quantum walks.
Finally, Sadowski \cite{sadowski04efficient} has recently provided an efficient
implementation of the quantum search algorithm exploiting the structure of
Apollonian networks.
\subsection*{Open quantum walks}\label{sec:oqw}
Following \cite{sinayskiy2013open} we recall the notion of Open Quantum Walks.
The model of the open quantum walk was introduced by Attal \emph{et al.}
\cite{attal12open} (see also \cite{sinayskiy12open}). In order to describe the
model, we consider a walk on a graph with the set of vertices $V$ and directed
edges $\{(i, j): \; i, j \in V\}$. The dynamics on the graph is described by
the space of states $\HH_2 = \mathbb{C}^V$ with the orthonormal basis $\{
\ket{i} \}_{i=0}^{|V| - 1}$. We describe an internal degree of freedom of the
walker by attaching a Hilbert space $\HH_1$ to each vertex of the graph. Hence,
the state of the quantum walker is described by the element of the space
$\LL(\HH_1 \otimes \HH_2)$.

Let us imagine a single quantum particle wandering through the vertices of a
graph. In discrete moments of time, the particle hops from one vertex $i$ to
another vertex $j$. With each transition the quantum state of the particle is
changed by a quantum operation associated with the edge $(i, j)$. With each step
the particle can, but does not have to, hop to all neighbours of vertex $i$.
Thus, after several steps the particle may become ``smeared'' over the vertices
of the graph.

\subsection*{Quantum states and quantum channels}\label{sec:quantu-channels}
In the following we recall the standard notions of quantum mechanics that are
essential for understanding the content of this paper.
\begin{definition}
    Linear Hermitian operator $\rho\in \LL(\HH)$ that is positive semi-definite
    ($\rho \geq 0$) and has a trace lesser or equal to one ($\Tr(\rho)\leq1$) is
    called a sub-normalized quantum state. A set of sub-normalized quantum states
    acting on $\HH$ will be denoted as $\Omega_\leq(\HH)$.
\end{definition}

\begin{definition}
    If a sub-normalized quantum state has a unit trace ($\Tr{\rho}=1$), then it
    is called a quantum state. A set of quantum states acting on $\HH$ will be
    denoted as $\Omega(\HH)$.
\end{definition}

\begin{definition}
    A linear map $\Phi: \LL(\HH_I) \rightarrow \LL(\HH_O)$ is completely
    positive (CP) iff for some $K$ it can be written as
    \begin{equation}
    \Phi(\rho)=\sum_{k=1}^K E_k^{\vphantom{\dagger}} \rho E_k^\dagger,
    \end{equation}
    where $E_k \in \LL(\HH_I, \HH_O)$ are called Kraus operators and $\rho\in \LL(\HH_I)$.
\end{definition}

\begin{definition}
    A linear map $\Phi: \LL(\HH_I) \rightarrow \LL(\HH_O)$ is trace non-increasing (TNI) iff 
    \begin{equation}
    \Tr(\Phi(\rho))\leq 1, \forall \rho\in\Omega(\HH_I).
    \end{equation}
\end{definition}

\begin{definition}
    A linear map $\Phi: \LL(\HH_I) \rightarrow \LL(\HH_O)$ is trace preserving
    (TP) iff 
    \begin{equation}
    \Tr(\Phi(\rho))= 1, \forall \rho\in\Omega(\HH_I).
    \end{equation}
\end{definition}
\begin{definition}
    A linear map $\Phi$ that is completely positive and trace non-increasing 
    (CP-TNI) is called a quantum operation.
\end{definition}

\begin{definition}
    A linear map $\Phi$ that is completely positive and trace preserving
    (CP-TP) is called a quantum channel.
\end{definition}

\begin{remark}
CP-TNI map given by Kraus operators $E_k \in \LL(\HH_I, \HH_O)$ fulfills the
condition $\sum_k E_k^\dagger E_k^{\vphantom{\dagger}}\leq\1_{\HH_O}$,
accordingly CP-TP fulfills the condition $\sum_k E_k^\dagger
E_k^{\vphantom{\dagger}} = \1_{\HH_O}$ \cite{bengtsson2006geometry}.
\end{remark}

\begin{definition}
    Mapping $\mu:O\to F$ from a~finite set of measurement outcomes
    $O=\{o_i\}_{i=1}^N$ into a set of measurement operators $F=\{A_i: A_i\in
    \LL(\HH_I, \HH_O)\}_{i=1}^N$ that fulfills the following relation
    \begin{equation}
        \sum_{i=1}^N A_i^\dagger A_i=\1_{\HH_O}
    \end{equation} 
    is called a quantum measurement. Probability $p_i$ of measuring the outcome
    $o_i$ in the state $\rho$ is given by $p_i=\Tr({A_i\rho A_i^\dagger})$.
    Given the measurement outcome $o_i$ the sub-normalized quantum state after
    the measurement $\mu$ is given by $\rho_{o_i} = A_i\rho
    A_i^\dagger$.
\end{definition}
In this work we limit ourselves to square projective orthonormal measurement
operators \textit{i.e.} $A_i \in \LL(\HH_I)$, $A_i^2=A_i$ for all $i \in
1,\ldots, N$ and for all $i,j\in 1\ldots,N$ $A_i A_j=\delta_{ij}A_i$.

\subsection*{Generalized open quantum walks}\label{sec:goqw}
To formally describe the dynamics of the generalized open quantum walk we
introduce a quantum operation $\TOM{E}_{ij}$, for each edge $(j, i)$. This
operation describes the change in the internal degree of freedom of the walker
due to the move from vertex $j$ to vertex $i$. We impose the limitation that the
sum of all quantum operations associated with the edges leaving vertex $j$ form
a quantum channel.

To describe generalized open quantum walks we use the notion of Transition
Operation Matrices, introduced in \cite{gudder2008quantum}, which provides a
generalization of stochastic matrices.

\begin{definition}
    Sub-Transition Operation Matrix (sub-TOM)
    $\TOM{E}=\{\mathcal{E}_{ij}\}_{i,j=1}^{M,N}$ is a matrix of
    completely-positive trace non-increasing (CP-TNI) maps such that
    \begin{equation}
    \forall_{1\leq j \leq N} \sum\limits_{i=1}^{M} \mathcal{E}_{ij} = \Phi_j, 
    \label{eq:outgoing}
    \end{equation}
    where $\Phi_j$ are completely positive trace non-increasing (CP-TNI) maps.
\end{definition}

\begin{definition}\label{def:TOM}
Transition Operation Matrix (TOM) is a sub-TOM with every $\Phi_j$ being a
completely positive trace preserving (CP-TP) map.
\end{definition}

In this work we will only consider square TOMs, therefore in what follows we
assume $M=N$. For the sake of simplicity we assume that all operators
$\mathcal{E}_{ij}: \LL(\HH_1) \rightarrow \LL(\HH_1)$ act on qudits of dimension
$\dim \HH_1$ and produce qudits of the same dimension.

\begin{remark}
    If $\TOM{E}=\{\mathcal{E}_{ij}\}_{i,j=1}^{M,K}$ and
    $\TOM{F}=\{\mathcal{F}_{ij}\}_{i,j=1}^{K,N}$ are TOMs, then their product
    $\TOM{G}=\TOM{E}\TOM{F}$ is also a TOM such that
    $\TOM{G}_{ij}=\sum\limits_{k=1}^{K}\TOM{E}_{ik}\TOM{F}_{kj}$
    \cite{gudder2008quantum}. Accordingly, a product of two sub-TOMs is also a
    sub-TOM.
\end{remark}

With TOM $\TOM{E}$ one can associate a Quantum Markov chain according to the
following definition.
\begin{definition}
Quantum Markov chain is a finite directed graph $G=(E,V)$ labelled by
$\TOM{E}_{ij}$ for $e\in E$ and by zero operator for $e \in \overline{E}$, with
$e\in V\times V$.
\end{definition}
Quantum Markov chain can be represented as $N\times N$ TOM, where $N=|V|$.
The state of quantum Markov chain is given by a vector state defined as follows.
\begin{definition}
    Sub-vector state is a column vector 
    $\alpha=(\alpha_1,\alpha_2,\ldots,\alpha_N)^T$ such
    that $\alpha_i$ are sub-normalized quantum states \textit{i.e.} 
    $\alpha_i\in\Omega_\leq(\HH)$, and
    $\sum_{i=1}^{N} \alpha_i\in\Omega_\leq(\HH)$ is a sub-normalised quantum state.
\end{definition}

\begin{definition}
    Vector state is a sub-vector state for which $\sum_{i=1}^{N}
    \alpha_i\in\Omega(\HH)$ is a quantum state.
\end{definition}

\begin{figure}[!h]
    \begin{center}
        \includegraphics{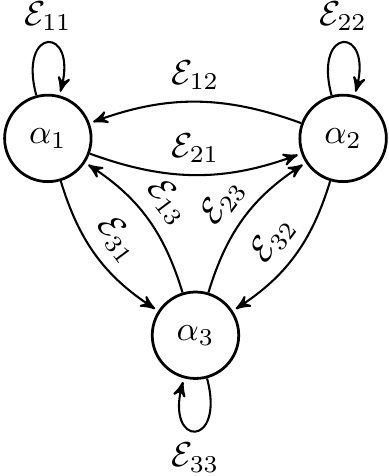}
    \end{center}
    \caption[]{
        {\bf An example of a three state TOM 
            $
            \TOM{E}=\left[
            \protect\begin{smallmatrix}
            \mathcal{E}_{11} & \mathcal{E}_{12} & \mathcal{E}_{13} \\
            \mathcal{E}_{21} & \mathcal{E}_{22} & \mathcal{E}_{23} \\
            \mathcal{E}_{31} & \mathcal{E}_{32} & \mathcal{E}_{33} 
            \protect\end{smallmatrix}
            \right]
            $.}
        Here $\alpha_i$-s in vertices denote sub-normalized quantum states associated
        with respective vertices at the given moment of time, therefore the state of the OQW
        can be described by a vector state $\alpha=(\alpha_{1}, \alpha_{2},
        \alpha_{3})^{T}$.
    }
    \label{fig:tom}
\end{figure}

Action $\TOM{E}(\alpha^{(t)})$ of (sub-)TOM $\TOM{E}$ on a (sub-)vector state
$\alpha^{(t)}$ at moment $t$ produces (sub-)vector state $\alpha^{(t+1)}$ at
moment $t+1$. This action is obtained in the following way:
$\alpha_i^{(t+1)}=\sum_{j=1}^{N}\TOM{E}_{ij}(\alpha_j^{(t)})$. An example of a
graph associated with a TOM is presented in Fig.~\ref{fig:tom}.

One should note that, in the case of one-dimensional internal state space,
$\dim\HH_1=1$, the operators $\mathcal{E}_{ij}$ become real numbers and form a 
stochastic matrix and thus the introduced chain is equivalent to the classical 
Markov chain.

\subsection*{TOMs as quantum channels}\label{sec:gen-oqw}
Let $\TOM{E}=\{\TOM{E}_{ij}\}_{i=1,j=1}^{N,N}$ be a TOM of dimensions $N\times
N$ with elements acting on $\LL(\HH_1)$. Let each of TOM's elements
$\TOM{E}_{ij}: \LL(\HH_1)\rightarrow\LL(\HH_1)$ have Kraus operators
$\{E_{k_{ij}}\}_{k_{ij}=1}^{K_{ij}}$, where $E_{k_{ij}}\in \LL(\HH_1)$ and
${K_{ij}}\in \mathbb{N}_+$, therefore the action of the elements is given by
$\TOM{E}_{ij}(\cdot)=\sum\limits_{k_{ij}=1}^{K_{ij}} E_{k_{ij}ij}\cdot
E_{k_{ij}ij}^\dagger$.

Let us construct the set of operators
$\{\hat{E}_{k_{ij}ij}\}_{k_{ij}=1,i=1,j=1}^{K_{ij},N,N}$
$\hat{E}_{k_{ij}ij}\in\LL(\HH_1\otimes\HH_2)$ in the form
$\hat{E}_{k_{ij}ij}=E_{k_{ij}ij}\otimes\ketbra{i}{j}$, where
$\{\ket{i}\}_{i=1}^{N}$ and $\{\ket{j}\}_{j=1}^{N}$ span computational
orthonormal bases in $\HH_1$.

\begin{definition}
A linear map $\Phi_\TOM{E}:\LL(\HH_1\otimes \HH_2)\rightarrow \LL(\HH_1\otimes 
\HH_2)$ associated with TOM $\TOM{E}$ is defined by the set of operators 
$\{\hat{E}_{k_{ij}ij}\}_{k_{ij}=1,i=1,j=1}^{K_{ij},N,N}$.
\end{definition}

In what follows we show that, if map $\Phi_\TOM{E}$ is associated with a TOM
$\TOM{E}$, then it is CP-TP.
\begin{proposition}
If a set of operators
$\{E_{k_{ij}ij}\}_{k_{ij}=1,i=1,j=1}^{K_{ij},N,N}$
forms a TOM then the set of operators
$\{\hat{E}_{k_{ij}ij}\}_{k_{ij}=1,i=1,j=1}^{K_{ij},N,N}$
forms a quantum channel.
\end{proposition}
\begin{theproof}
To prove this claim it is sufficient to show that operators 
$\hat{E}_{k_{ij}}$ fulfill the completeness relation.
\begin{eqnarray*}
\sum\limits_{i=1}^{N}
\sum\limits_{j=1}^{N}
\sum\limits_{k=1}^{K_{ij}}
\hat{E}_{k_{ij}ij}^\dagger \hat{E}_{k_{ij}ij}&=&
\sum\limits_{i=1}^{N}
\sum\limits_{j=1}^{N}
\sum\limits_{k=1}^{K_{ij}}
(E_{k_{ij}ij}\otimes\ketbra{i}{j})^\dagger (E_{k_{ij}ij}\otimes\ketbra{i}{j}) 
\nonumber \\
&=&
\sum\limits_{i=1}^{N}
\sum\limits_{j=1}^{N}
\sum\limits_{k=1}^{K_{ij}}
E_{k_{ij}ij}^\dagger E_{k_{ij}ij}\otimes\ketbra{j}{j} \nonumber \\
&=&
\sum\limits_{j=1}^{N}
\left(
\sum\limits_{i=1}^{N}
\sum\limits_{k=1}^{K_{ij}}
E_{k_{ij}ij}^\dagger E_{k_{ij}ij}
\right)
\otimes\ketbra{j}{j} \nonumber \\
&=&
\sum\limits_{j=1}^{N} \1_{\HH_1} \otimes \ketbra{j}{j}\nonumber \\
&=&
\1_{\HH_1\otimes\HH_2}.
\end{eqnarray*}
\end{theproof}

\begin{theorem}
Let $\alpha=(\alpha_1,\ldots,\alpha_j,\ldots,\alpha_N)^T$ be a vector state.
With $\alpha$ we associate a block diagonal quantum state
\begin{equation}
\rho_\alpha=\sum\limits_{j=1}^{N}\alpha_j\otimes\ketbra{j}{j}\in 
\Omega(\HH_1\otimes \HH_2),
\end{equation}
where $N=\dim\HH_2$.
Accordingly let 
$\beta=(\beta_1,\ldots,\beta_i,\ldots,\beta_N)^T$ be a vector state with an 
associated state
\begin{equation}
\rho_\beta=\sum\limits_{i=1}^{N}\beta_i\otimes\ketbra{i}{i}\in 
\Omega(\HH_1\otimes \HH_2).
\end{equation}

Let $\Phi_\TOM{E}$ be a quantum channel associated with TOM $\TOM{E}$ and $\beta = \TOM{E}(\alpha)$, then
\begin{equation}
\rho_\beta=\Phi_\TOM{E}(\rho_\alpha).
\end{equation}
\end{theorem}
\begin{theproof}
\begin{eqnarray*}
\Phi_\TOM{E}(\rho_\alpha) 
&=&
\sum\limits_{i=1}^{N}
\sum\limits_{j=1}^{N}
\sum\limits_{k_{ij}=1}^{K_{ij}}
\hat{E}_{k_{ij}ij}\rho_\alpha \hat{E}_{k_{ij}ij}^\dagger \nonumber \\
&=&
\sum\limits_{i=1}^{N}
\sum\limits_{j=1}^{N}
\sum\limits_{k_{ij}=1}^{K_{ij}} 
(E_{k_{ij}ij}\otimes\ketbra{i}{j}) 
(\alpha_j \otimes \ketbra{j}{j})
(E_{k_{ij}ij}\otimes\ketbra{i}{j})^\dagger 
\nonumber \\
&=&
\sum\limits_{i=1}^{N}
\sum\limits_{j=1}^{N}
\sum\limits_{k_{ij}=1}^{K_{ij}} 
E_{k_{ij}ij}
\alpha_j
E_{k_{ij}ij}^\dagger \otimes\ketbra{i}{i}\nonumber\\
&=&
\sum\limits_{i=1}^{N}
\sum\limits_{j=1}^{N}
\TOM{E}_{ij}(\alpha_j)\otimes\ketbra{i}{i}=
\sum\limits_{i=1}^{N}
\beta_i\otimes\ketbra{i}{i}\nonumber\\
&=&\rho_\beta
\end{eqnarray*}
\end{theproof}

\begin{remark}
Generalized open quantum walks coincide with open
quantum walks introduced in \cite{attal12open} if all the operations have
Kraus rank equal to one \text{i.e.} can be described with a single Kraus
operator.
\end{remark}

\subsection*{Mean first passage time}\label{sec:hitting}
There is a number of random walk properties that one can consider in order to
analyse walk behaviour. In this paper we focus on the properties commonly
examined in the case of classical homogeneous random walks \ie\ walks with
transition probabilities evenly distributed and equal to $1/d$ for each vertex
of degree $d$. In particular, we study the mean first passage time and the
average return time. The former one describes the average time it takes to make
a move between two fixed nodes.
\begin{definition}
The mean first passage time (MFPT) from vertex $i$ to vertex $j$ is defined as 
the average time for the walker to reach vertex $j$ starting from vertex $i$:
\begin{equation}
T_{ij} = \sum_{t=1}^\infty t P_{ij}(t),
\end{equation}
where $P_{ij}(t)$ is the first passage probability from $i$ to $j$ after time 
$t$.
\end{definition}
\begin{definition}
The average return time (ART) $T_{ii}$ is the mean first passage time from 
vertex $i$ to itself.
\end{definition}

It has been shown that, in the case of homogeneous classical random walks, the
ART does not depend on the structure of the network, but only on the degree of
the vertex~\cite{noh2004random}. More precisely, the ART in this case is given
by:
\begin{equation}
T_{ii} = \frac{\sum_{j=1}^N d_j}{d_i},
\end{equation}
where $d_j$ denotes the degree of the $j$\textsuperscript{th} vertex. On the
other hand, the MFPT depends on the structure of the network and $T_{ij}$ do not
need be equal to $T_{ji}$.

\subsection*{Quantum mean first passage time}\label{sec:hitting}

The classical notion of reaching a vertex does not have an appropriate quantum
counterpart. There are some subtleties that make defining a quantum analogue a
troublesome task. The main difficulty lies in the measurement problem.

Let us recall the picture with a single quantum particle wandering through a
graph. Now we can imagine that we have placed a measurement apparatus at each
vertex of the graph. This apparatus performs an arbitrary quantum measurement
$\mu$. If the quantum measurement is trivial, \textit{i.e.} $\mu(o) =
\1_{\HH_1}$, then it allows only to check whether the particle is placed in a
given vertex. We can also choose a measurement that would tell us if the
particle has a given property. This is done using a measurement having two
values: $\Pi_v^{\phantom{\perp}}, \Pi_v^\perp = \1_{\HH_1} - \Pi_v$. Hereafter,
we will call $\Pi_v$ the view operator.

Let us construct the following sub-TOM $\TOM{F}$ based on a given TOM $\TOM{E}$
such that:
\begin{equation}
\TOM{F} = \TOM{E}\TOM{P},
\end{equation}
where $\mathcal{P}$ is diagonal sub-TOM with identity operators on the diagonal
except element $P_{jj}$, where $P_{jj}(\cdot)=\Pi_v^\perp \cdot 
\Pi_v^\perp$.
\begin{definition}\label{def:mfpt}
The quantum Mean First Passage Time (qMFPT) of $(\mathcal{E}, V, \rho_0, \Pi_v,
i, j)$, where $\mathcal{E}$ is a TOM, $\rho_0 \in \Omega(\HH_1)$ is a quantum
state, $\Pi_v$ is a view operator and $i, j \in V$ is:
\begin{equation}
Q_{ij} = \sum_{t=1}^{\infty} \Tr(\Pi_v \alpha_j^{(t)})t,\label{eq:qmfpt}
\end{equation}
$\alpha^{(t)}$ is given by:
\begin{equation}
\alpha^{(t)} = \TOM{F}(\alpha^{(t-1)}),
\end{equation}
and $\alpha^{(0)}$ is a state vector with $\rho_0$ at the $i$-th element and 
all other elements equal to zero.
\end{definition}

\begin{remark}
When $\dim \HH_1 = 1$, the open quantum walk reduces to a classical random 
walk. Therefore the introduced notion of qMFPT reduces to the classical case as 
well.
\end{remark}

\begin{definition}
The vertex-qMFPT of $(\mathcal{E}, V, \rho_0, \Pi_v, j)$ is:
\begin{equation}
Q_j = \frac{\sum_{i=1, i \neq j} ^ {|V|} Q_{ij}}{|V| - 
1},\label{eq:vertex-qmfpt}
\end{equation}
where $i, j \in V$.
\end{definition}

\begin{definition}
The degree-qMFPT of $(\mathcal{E}, V, \rho_0, \Pi_v, d)$, where $d \in 
\mathbb{N}$ is:
\begin{equation}
Q^{(d)} = \frac{\sum_{i \in V_d} Q_i}{|V_d|},\label{eq:degree-qmfpt}
\end{equation}
where $V_d \subset V$ is the set of all vertices with degree $d$.
\end{definition}

\begin{definition}
The degree-qART of $(\mathcal{E}, V, \rho_0, \Pi_v, d)$, where $d \in 
\mathbb{N}$ is:
\begin{equation}
Q_\mathrm{ART}^{(d)} = \frac{\sum_{i \in V_d} 
Q_{ii}}{|V_d|},\label{eq:degree-qart}
\end{equation}
where $V_d \subset V$ is the set of all vertices with degree $d$.
\end{definition}

\begin{remark}
In an Apollonian network, vertices of a given generation have equal degrees.
\end{remark}

\begin{remark}
By $T_j$, $T^{(d)}$ and $T_\mathrm{ART}^{(d)}$ we denote the classical 
vertex-MFPT, degree-MFPT and degree-ART, respectively.
\end{remark}

\begin{remark}
Calculating analytically the qMFPT as shown in Definition~\ref{def:mfpt} is a 
complicated problem. Thus, we turned to numerical simulations to obtain 
approximate results for a selected number of test cases. In order to 
numerically calculate the qMFPT, we have to limit $t$ in Eq~\eqref{eq:qmfpt} to 
a finite value $t_\mathrm{s}$. We choose such a value that:
\begin{equation}
1 - \sum_{t=1}^{t_\mathrm{s}} \Tr(\Pi_v \alpha_j^{(t)}) < 10^{-6}.
\end{equation}
\end{remark}
\section*{Discussion}\label{sec:examples}
Let us now focus our attention on the quantumness of the model introduced in the
previous section. In the following sections we provide a number of examples that
allow the observation of non-classical phenomena. However, it is not always the
case. It is crucial to note that, with appropriate TOM design, the walk mimics a
classical one.

\begin{remark}\label{rmk:unitary-is-classical}
For any open quantum walk designed with the use of unitary transformations
exclusively the position probability distribution at each step of the walk is
identical to the classical counterpart.

Moreover, the values of qMFPT and qART match the classical counterparts when the
identity $\1$ view operator is considered regardless of the initial state.
\end{remark}

Let us introduce a simple walk that allows tracking its evolution in detail in
order to provide an example of non-classical behaviour.

\subsection*{Simple example}
\begin{figure}[!h]
\centering
\includegraphics{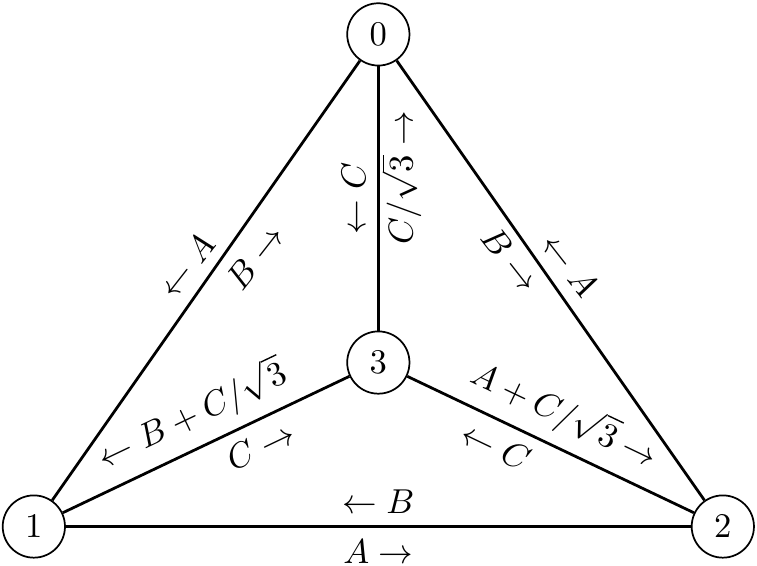}
\caption{{\bf Apollonian network with 4 vertices.} Operators $A$, $B$ and $C$ are 
defined in the text.}\label{fig:network-4-nodes}
\end{figure}

As the first example we study the four-vertex Apollonian network with a walking
qutrit. This network is schematically depicted in
Fig.~\ref{fig:network-4-nodes}. We study three different view operators:
$
A = \ketbra{x}{x},
B = \ketbra{y}{y},
C = \ketbra{z}{z},
$
where $A, B, C \in \LL(\mathbb{C}^3)$ and
\begin{alignat}{3}
\ket{x} &= [1 \; 1 \; 1]^\mathrm{T} / \sqrt{3},&\quad
\ket{y} &= [1 \; \omega \; \omega^2]^\mathrm{T} / \sqrt{3},&\quad
\ket{z} &= [1 \; \omega^2 \; \omega^4]^\mathrm{T} / \sqrt{3},
\end{alignat}
with $\omega=e^{2\pi  \ii / 3}$.
We choose the initial state of the walker to be the maximally mixed state
localized at the central vertex
\begin{equation}
\alpha^{(0)} = (0_{\mathbb{C}^3}, 0_{\mathbb{C}^3}, 0_{\mathbb{C}^3}, \frac13 
\1_{\mathbb{C}^3})^\mathrm{T}.
\end{equation}

\begin{figure}[!h]
\centering
\includegraphics{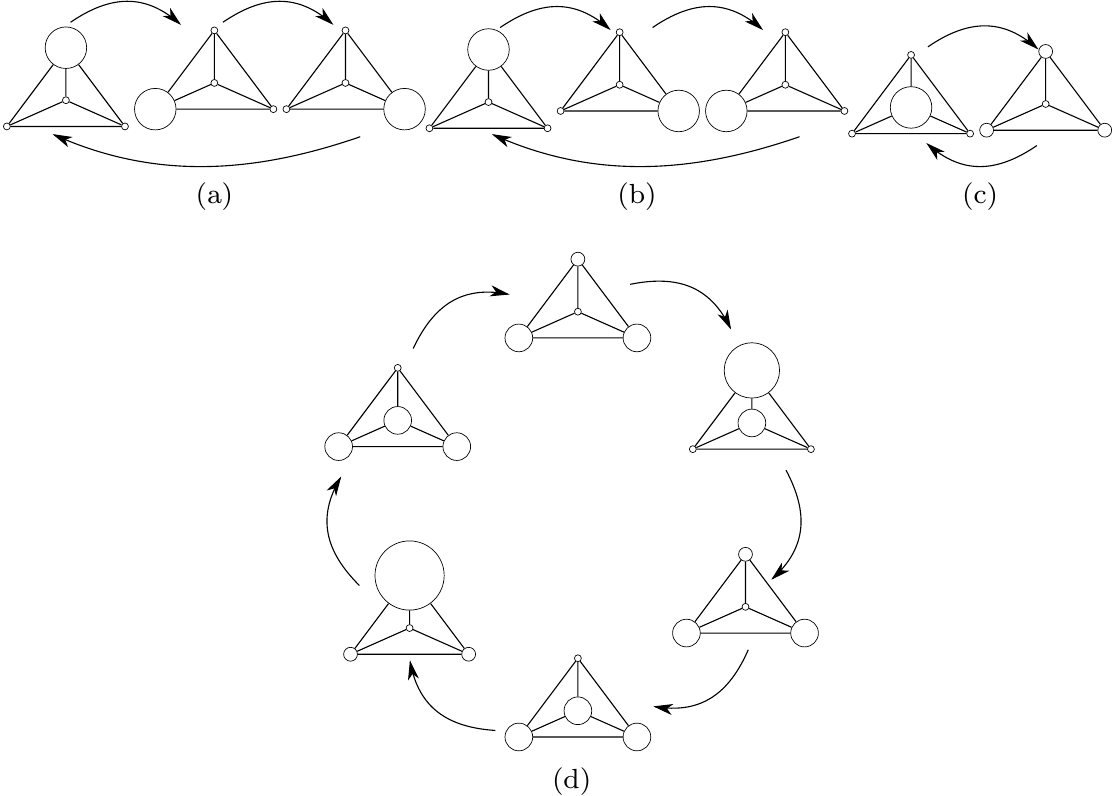}
\caption{{\bf Open quantum walk on an Apollonian network with 4 vertices.} Each
panel of this Figure shows the behaviour of the network in subspaces associated
with selected measurement operators. The size of the vertices is proportional
to the probability of measuring the walker in that vertex. The picture
represents the evolution after the first step of the walk. Panel (a) corresponds
to $\Pi_\mathrm{v}=A$, panel (b) to $\Pi_\mathrm{v}=B$, panel (c) to
$\Pi_\mathrm{v}=C$ and panel (d) to $\Pi_\mathrm{v}=\1$.
}\label{fig:walk-4-nodes}
\end{figure}

The probability distributions of measuring the particle depending on the view
operators are depicted in Fig.~\ref{fig:walk-4-nodes}. Note that in the initial
state the particle is located in vertex three. After the first step the
behaviour becomes cyclic. Fig.~\ref{fig:walk-4-nodes}a shows the behaviour of
the walker in the subspace associated with view operator~$A$. Accordingly,
Fig.~\ref{fig:walk-4-nodes}b shows the same result in the case of view
operator~$B$ and Fig.~\ref{fig:walk-4-nodes}c for the operator~$C$. The complete
behaviour of the walker is shown in Fig.~\ref{fig:walk-4-nodes}d. In the first
subspace we achieved a counter-clockwise walk on the external vertices. In the
second subspace, we achieved a clockwise walk on the external vertices. Both of
these walks have a period $T=3$. In the third subspace we achieved an
oscillating behaviour, between the central and external vertices so the walk has
a period of $T=2$. Thus, the entire walk is periodic with period $T=6$. Hence we
have shown that the evolution of the open quantum walk heavily depends on the
view operator.

The explicit form of TOM depicted in Fig.~\ref{fig:network-4-nodes} reads 
\begin{equation}
\mathcal{E} = \left[
\begin{array}{cccc}
0_{\LL(\mathbb{C}^3)} & B \cdot B^\dagger  & A \cdot A^\dagger & \frac13 C 
\cdot 
C^\dagger \\
A \cdot A^\dagger & 0_{\LL(\mathbb{C}^3)} & B \cdot B^\dagger & (B + 
C/\sqrt{3}) 
\cdot (B + C/\sqrt{3})^\dagger \\
B \cdot B^\dagger & A \cdot A^\dagger & 0_{\LL(\mathbb{C}^3)} & (A + 
C/\sqrt{3}) \cdot (A + C/\sqrt{3})^\dagger \\
C \cdot C^\dagger & C \cdot C^\dagger & C \cdot C^\dagger & 
0_{\LL(\mathbb{C}^3)}
\end{array}
\right],
\end{equation}
where $\mathcal{E}_{ij}(\cdot) = X \cdot X^\dagger$ denotes a rank-one quantum
operation. Using $\mathcal{E}$ and setting $V=\{0, 1, 2, 3\}$, $\rho_0 = \frac13
\1_{\mathbb{C}^3}$ and $\Pi_v$ equal to $\1_{\mathbb{C}^3}$, $\ketbra{0}{0}$,
$\ketbra{1}{1}$, $\ketbra{2}{2}$, $A$, $B$ or $C$ we compute the qMFPTs and
qARTs of this open quantum walk as shown in Tab.~\ref{tab:mfpt-4-nodes}. Value
$\infty$ means that the state cannot be reached from a given initial state. The
diagonal entries are the qARTs. For comparison in
Tab.~\ref{tab:mfpt-4-nodes-classical} we show the MFPTs and ARTs in the
classical homogeneous random walks on this graph. The ``quantumness'' --- the
non-classical behaviour --- of the walk can be seen in the view-conditioned
qMFPTs. 

The walk designed in this example is based on a non-bistochastic transition
operators~\cite{bengtsson2006geometry}. This allows us to demonstrate the sharp
non-classical behaviour of the walk that can occur in such case.
\section*{Experiments}

The discussion above gives some insight on the relation between walk behaviour and
its crucial properties such as the TOM design, view operator form and the state
chosen to define the starting conditions. Now we discuss a series of experiments
that illustrate the possibility of obtaining non-classical behaviour in
generalized open quantum walks considering these properties.

We will rate the walk quantumness in the terms of the qMFPT and qART by analysing
the following cases:
\begin{itemize}
\item nearly classical walk where the quantum behaviour is initial state 
dependent,
\item a walk based on a classical TOM for which the view operator determines 
the observed walk properties,
\item an open quantum walk that exhibits strong non-classical phenomena for any 
view operator.
\end{itemize}

\subsubsection*{Case 1 -- quantum counterpart}
In this example we aim at constructing an OQW which resembles a classical random
walk, differing from it only in some minor features. In order to achieve this,
let us consider an open quantum walk which by construction mimics the structure
of classical homogeneous random walk on an Apollonian network. By a homogeneous
walk we understand a random walk for which the exit probability from every
vertex in any allowed direction is equal to one over the degree of the vertex. 
In this example we will set the space associated with the integral degree of 
freedom of the walker to be a qutrit space, i. e. $\HH_1 = \mathbb{C}^3$.
The walk is constructed in the following way:
\begin{itemize}
\item For every edge outgoing from vertices in the last generation of the
Apollonian network, we assign TOM elements with two associated Kraus operators
given by rescaled projections on two mutually orthogonal subspaces so that
condition~in Def.~\ref{def:TOM} holds.
\item Every other TOM element is a rescaled identity operator.
\end{itemize}

Specifically, each outgoing edge of 
the vertices in the last generation is assigned one of the following pairs of 
Kraus operators:
\begin{equation}
\begin{split}
\mathcal{P}_1 = \left(\frac{1}{\sqrt{2}} \ketbra{0}{0}, \frac{1}{\sqrt{2}} 
\ketbra{1}{1} 
\right),\\
\mathcal{P}_2 = \left(\frac{1}{\sqrt{2}} \ketbra{1}{1}, \frac{1}{\sqrt{2}} 
\ketbra{2}{2} 
\right),\\
\mathcal{P}_3 = \left(\frac{1}{\sqrt{2}} \ketbra{2}{2}, \frac{1}{\sqrt{2}} 
\ketbra{0}{0} 
\right).
\end{split}\label{eq:pairs}
\end{equation}
As each vertex in the last generation has degree $d=3$, it is easily seen that 
this assignment fulfills Def.~\ref{def:TOM}.

For the sake of clarity we will show the behavior of the walks on a
3\textsuperscript{rd} generation Apollonian network. In this case, we can write
the assignment of pairs~\eqref{eq:pairs} explicitly:
\begin{equation}
\begin{split}
E_{1,0,7} = \ketbra{0}{0} / \sqrt{2}, \quad  E_{2,0,7} = \ketbra{1}{1} / 
\sqrt{2},& \quad
E_{1,1,7} = \ketbra{1}{1} / \sqrt{2}, \quad  E_{2,1,7} = \ketbra{2}{2} / 
\sqrt{2}, \\
E_{1,4,7} = \ketbra{2}{2} / \sqrt{2}, \quad  E_{2,4,7} = \ketbra{0}{0} / 
\sqrt{2},& \\
E_{1,0,8} = \ketbra{0}{0} / \sqrt{2}, \quad  E_{2,0,8} = \ketbra{1}{1} / 
\sqrt{2},& \quad
E_{1,4,8} = \ketbra{1}{1} / \sqrt{2}, \quad  E_{2,4,8} = \ketbra{2}{2} / 
\sqrt{2}, \\
E_{1,3,8} = \ketbra{2}{2} / \sqrt{2}, \quad  E_{2,3,8} = \ketbra{0}{0} / 
\sqrt{2},& \\
E_{1,1,9} = \ketbra{0}{0} / \sqrt{2}, \quad  E_{2,1,9} = \ketbra{1}{1} / 
\sqrt{2},& \quad
E_{1,4,9} = \ketbra{1}{1} / \sqrt{2}, \quad  E_{2,4,9} = \ketbra{2}{2} / 
\sqrt{2}, \\
E_{1,3,9} = \ketbra{2}{2} / \sqrt{2}, \quad  E_{2,3,9} = \ketbra{0}{0} / 
\sqrt{2},& \\
E_{1,1,10} = \ketbra{0}{0} / \sqrt{2}, \quad  E_{2,1,10} = \ketbra{1}{1} / 
\sqrt{2},&\quad
E_{1,2,10} = \ketbra{1}{1} / \sqrt{2}, \quad  E_{2,2,10} = \ketbra{2}{2} / 
\sqrt{2}, \\
E_{1,5,10} = \ketbra{2}{2} / \sqrt{2}, \quad  E_{2,5,10} = \ketbra{0}{0} / 
\sqrt{2},& \\
E_{1,1,11} = \ketbra{0}{0} / \sqrt{2}, \quad  E_{2,1,11} = \ketbra{1}{1} / 
\sqrt{2},&\quad
E_{1,5,11} = \ketbra{1}{1} / \sqrt{2}, \quad  E_{2,5,11} = \ketbra{2}{2} / 
\sqrt{2}, \\
E_{1,3,11} = \ketbra{2}{2} / \sqrt{2}, \quad  E_{2,3,11} = \ketbra{0}{0} / 
\sqrt{2},& \\
E_{1,2,12} = \ketbra{0}{0} / \sqrt{2}, \quad  E_{2,2,12} = \ketbra{1}{1} / 
\sqrt{2},&\quad
E_{1,5,12} = \ketbra{1}{1} / \sqrt{2}, \quad  E_{2,5,12} = \ketbra{2}{2} / 
\sqrt{2}, \\
E_{1,3,12} = \ketbra{2}{2} / \sqrt{2}, \quad  E_{2,3,12} = \ketbra{0}{0} / 
\sqrt{2},& \\
E_{1,0,13} = \ketbra{0}{0} / \sqrt{2}, \quad  E_{2,0,13} = \ketbra{1}{1} / 
\sqrt{2},&\quad
E_{1,2,13} = \ketbra{1}{1} / \sqrt{2}, \quad  E_{2,2,13} = \ketbra{2}{2} / 
\sqrt{2}, \\
E_{1,6,13} = \ketbra{2}{2} / \sqrt{2}, \quad  E_{2,6,13} = \ketbra{0}{0} / 
\sqrt{2},& \\
E_{1,2,14} = \ketbra{0}{0} / \sqrt{2}, \quad  E_{2,2,14} = \ketbra{1}{1} / 
\sqrt{2},&\quad
E_{1,6,14} = \ketbra{1}{1} / \sqrt{2}, \quad  E_{2,6,14} = \ketbra{2}{2} / 
\sqrt{2}, \\
E_{1,3,14} = \ketbra{2}{2} / \sqrt{2}, \quad  E_{2,3,14} = \ketbra{0}{0} / 
\sqrt{2},& \\
E_{1,0,15} = \ketbra{0}{0} / \sqrt{2}, \quad  E_{2,0,15} = \ketbra{1}{1} / 
\sqrt{2},&\quad
E_{1,6,15} = \ketbra{1}{1} / \sqrt{2}, \quad  E_{2,6,15} = \ketbra{2}{2} / 
\sqrt{2}, \\
E_{1,3,15} = \ketbra{2}{2} / \sqrt{2}, \quad  E_{2,3,15} = \ketbra{0}{0} / 
\sqrt{2}.& \\
\end{split}
\end{equation}
The numbering of vertices follows the convention shown in
Fig.~\ref{fig:example-network}.

In this case we use the following initial state for calculating degree-qMFPT and
degree-qART: $\rho_0=\ketbra{x}{x}$, where $\ket{x} = [1 \; 1 \; 1]^\mathrm{T} /
\sqrt{3}$. For such an initial state we obtain the results differing from
classical ones even when the view operator is equal to the identity
$\Pi_v=\1_{\HH_1}$. The resulting degree-qMFPTs and degree-qARTs are shown in
Tab.~\ref{tab:almost-classical}. The most significant difference from the
classical set-up is that the degree-qARTs do not scale as $\frac{1}{d}$, where
$d$ is the degree of the vertex. It should be noted that by the construction of
the model, when the initial state is a classical mixture $\rho_0 = \frac13
\1_{\mathbb{C}^3}$, we can recover the classical behaviour.

In the case of a walk defined mostly with the use of the identity operators the
vast majority of the transition operations is bi-stochastic. The disturbance is
introduced in the last generation nodes. We have shown that with a slight
change of the walk structure and with an appropriate initial state we observe a
significant alteration of the walk and non-classical behaviour, even without
considering the view operator.

\subsubsection*{Case 2 -- measurement manipulation}
In this case we aim to analyse the behaviour of a walk based on bi-stochastic
operations exclusively. In particular we investigate the behaviour in terms of
qMFTP and qART values when a variety of view operators is applied. We consider
an open quantum walk on the fifth generation Apollonian network, constructed as
follows.

Let $d_i$ be the degree of vertex $i$, the internal state space to be
two-dimensional $\HH_1=\mathbb{C}^2$, $G_1$, $G_2$ with $G_1 > G_2$ denote two
different generations of the Apollonian networks, $V_{G_1}$ and $V_{G_2}$ be the
sets of vertices in generations $G_1$ and $G_2$, respectively. By $i$ and $j$ we
denote vertices in generations $G_1$ and $G_2$, respectively, \textit{i.e.} $i
\in V_{G_1}$, $j \in V_{G_2}$. Then:
\begin{itemize}
\item For transitions from generation $G_1$ to generation $G_2$,we choose TOM 
elements equal to $\mathcal{E}_{ji}(\rho) =\frac1d \sigma_x \rho \sigma_x$.
\item For transitions from generation $G_2$ to generation $G_1$ we choose TOM 
elements equal to $\mathcal{E}_{ij}(\rho) = \frac1d \sigma_z \rho \sigma_z$.
\item In the case of the zeroth generations there exist intra-generation 
transitions. Let us denote by $k, l \in V_{G_0}$ the vertices in this 
generation. For these transitions we assign a rescaled identity operator 
$\mathcal{E}_{kl}(\rho) = \frac{1}{d}\rho$.
\end{itemize}
Here, $\sigma_x$ and $\sigma_z$ are the Pauli matrices given by:
\begin{equation}
\sigma_x = 
\begin{pmatrix}
0 & 1 \\
1 & 0
\end{pmatrix}, \quad
\sigma_z = 
\begin{pmatrix}
1 & 0 \\
0 & -1
\end{pmatrix}.
\end{equation}
We use the following view operators $\Pi_v$:
$\1_{\mathbb{C}^2}$, $\ketbra{0}{0}$,
$\ketbra{+}{+}$ and $\ketbra{j}{j}$,
with:
\begin{equation}
\begin{split}
\ket{+} =& \frac{1}{\sqrt{2}} \left( \ket{0} + \ket{1} \right), \\
\ket{j} =& \frac{1}{\sqrt{2}} \left( \ket{0} + \ii\ket{1} \right).
\end{split}
\end{equation}
In this case we choose the initial state to be $\rho_0 = \frac12
\1_{\mathbb{C}^2}$.

\begin{figure}[ht!]
\centering
\includegraphics[width=0.8\textwidth]{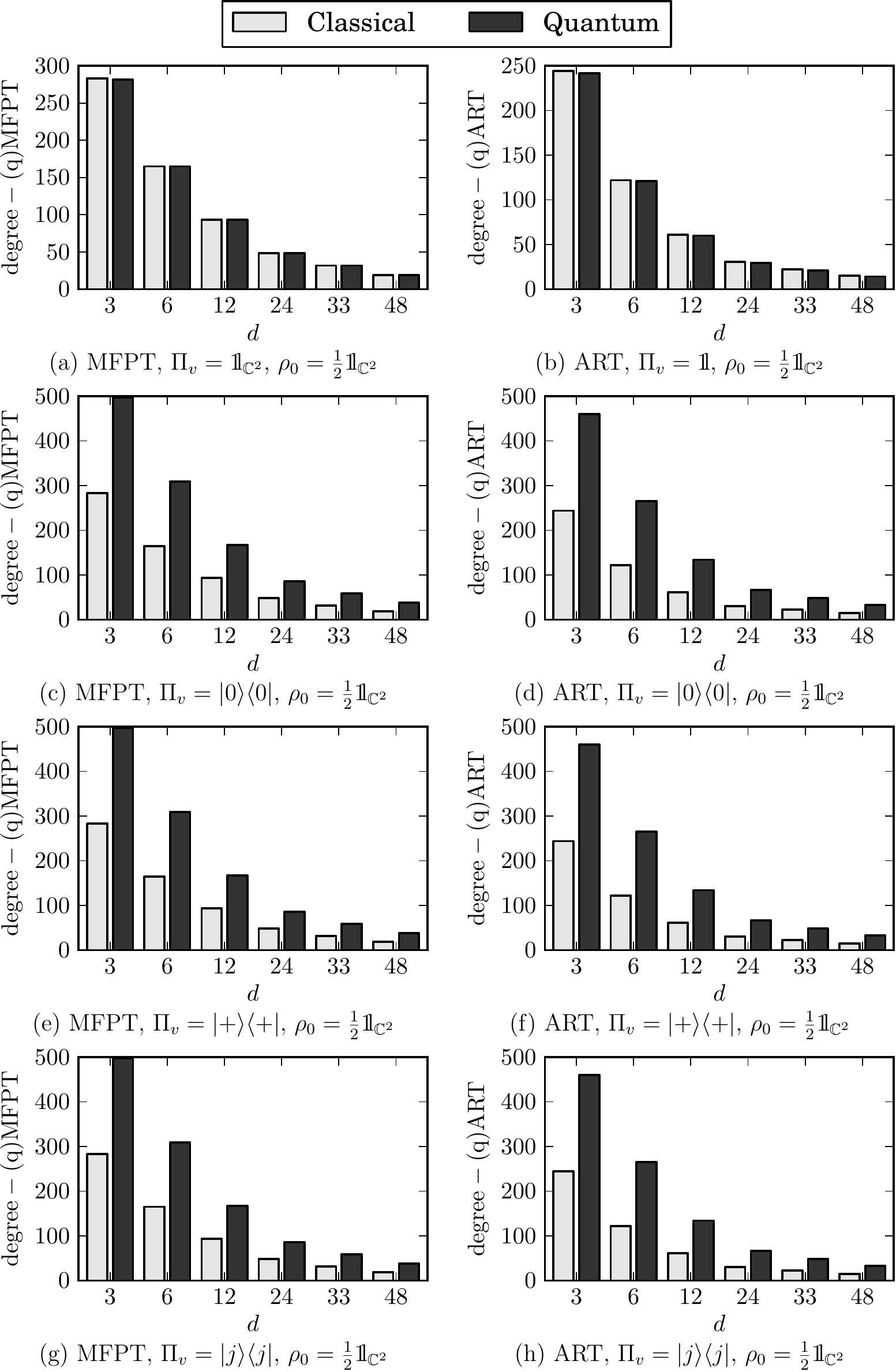}
\caption{{\bf Degree-(q)MFPTs and degree-(q)ARTs conditioned on the view
operator.} The labels denote the conditioning measurement
operators.}\label{fig:MFPT-generation}
\end{figure}

The results for this case are shown in Fig.~\ref{fig:MFPT-generation}. As all of
the transition operators are bi-stochastic, the walk exhibits exactly classical
behaviour when the view operator equals to $\1_{\mathbb{C}^2}$. Although the
view operators increase the value of degree-qMFPT, the values differ only by a
constant factor. Hence the overall trend remains unchanged. As in the previous
case, the main difference between the classical and quantum set-ups lies in the
degree-qARTs. Again, they do not scale as $\frac{1}{d}$. Thus, the view operator
is the key ingredient that allows the observation of non-classical behaviour in
the case of a walk with bi-stochastic transition operations.

\subsubsection*{Case 3 -- quantum effect}
In this case we again consider non-bistochastic walk. The walk does not mimic a
classical one and thus we are able to obtain striking differences in terms of
MFPT/ART behaviour. In this case the transition operator assignment is also
based on the generation of a vertex. More precisely, we divide vertices in the
graph into classes. Each class is identified by the set of generations of the
neighbouring vertices. As a result, each class corresponds to the vertices with
identical configuration of generations of neighbouring vertices.

\begin{figure}[!ht]
\centering
\includegraphics[width=0.6\textwidth]{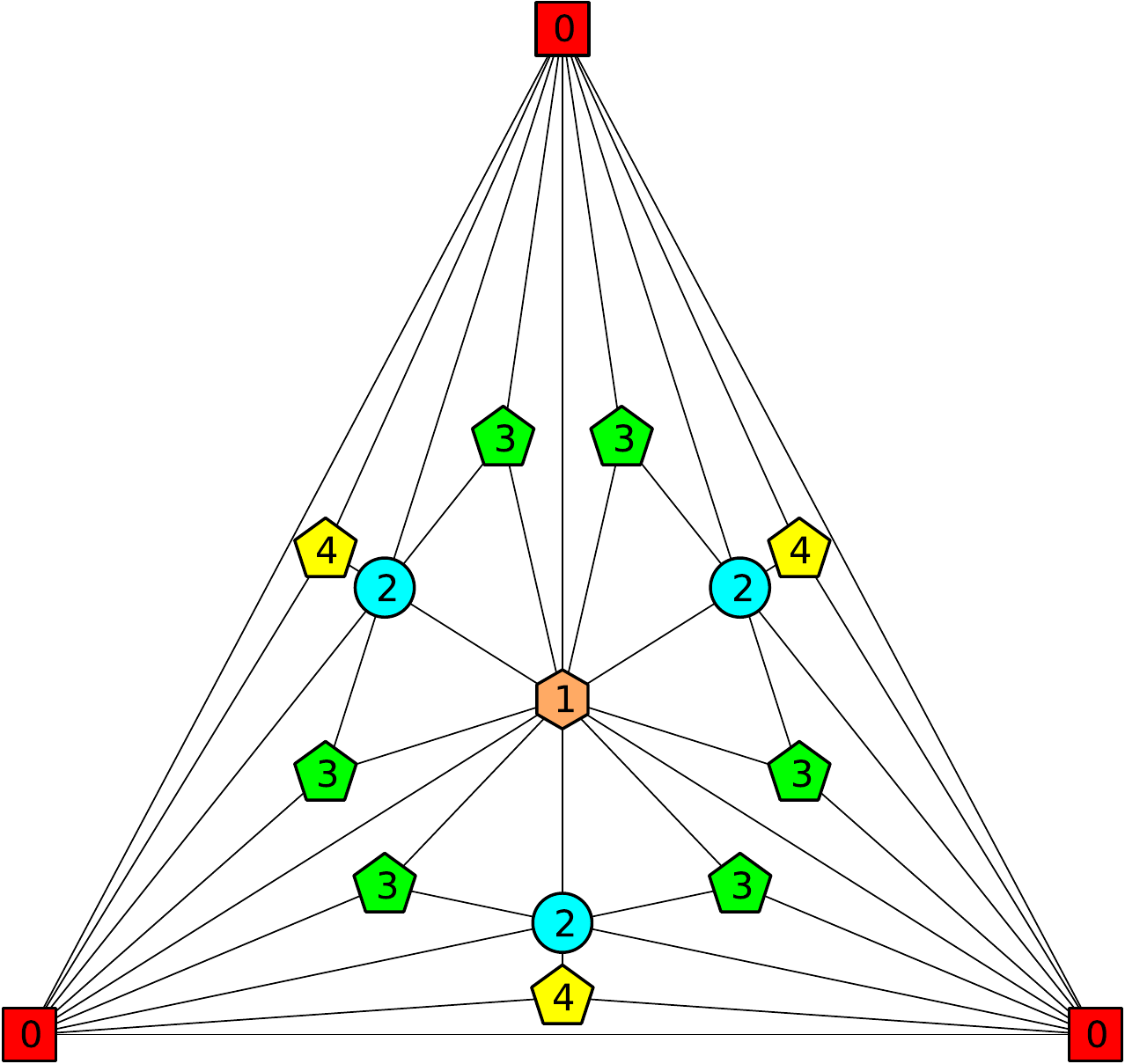}
\caption{{\bf The Apollonian network with 16 vertices (3\textsuperscript{rd}
generation) divided into 5 classes.} The classes are chosen based on the 
vertex generation and the generations of its neighbours. In this case, the 
green pentagons and yellow pentagons are of the same generation, but belong to 
different classes. The numbers denote the classes used in 
Eq.~\eqref{eq:operators-assignment}.}
\label{fig:network-classes}
\end{figure}

Here we consider the 3\textsuperscript{rd} generation of Apollonian network
consisting of 16 vertices divided into 5 classes as shown in
Fig.~\ref{fig:network-classes}. This approach allows us the simplified
assignment of operators, as the number of classes is significantly lower than
the number of vertices and provides strong symmetry of the network dynamics.

In order to design system dynamics we introduce two decompositions of the space
$\HH_1$ into mutually orthogonal subspaces. For each subspace we choose an
operator that acts on this subspace exclusively. In this example, we set
$\HH_1=\mathbb{C}^4$. We study two decompositions ($x$ and $z$) of $\HH_1$ with
the following operators:
\begin{equation}
B_x = (\1_{\mathbb{C}^4} - \sigma_x \otimes \sigma_x)/2,  \quad
C_x = (\1_{\mathbb{C}^4} + \sigma_x \otimes \sigma_x)/2,
\end{equation}
and
\begin{equation}
B_z = (\1_{\mathbb{C}^4} - \sigma_z \otimes \sigma_z)/2, \quad
C_z = (\1_{\mathbb{C}^4} + \sigma_z \otimes \sigma_z)/2.
\end{equation}

For every possible pair of classes $(c_1, c_2)$ we choose a set (with one or two
elements in this case) of Kraus operators $\{A^{(c_1, c_2)}_1, \ldots, A^{(c_1,
c_2)}_{n_{c_1, c_2}}\}$ from $\{B_x, C_x, B_z, C_z\}$. In order to satisfy
Def.~\ref{def:TOM}, we design transition operators assignment with two
normalization rules. When the designed assignment causes that, for some vertex
in the network, there are ($k$) outgoing edges assigned with the same operator
all these operators are multiplied by the factor $1/\sqrt{k}$. For example each
vertex of the class 1 has 6 outgoing edges corresponding to the $C_z$ operator:
3 neighbours of class 0 and 3 neighbours of class 2. Thus we introduce
$1/\sqrt{6}$ normalization factor for the $A_1^{(1,0)}$ and $A_1^{(1,2)}$
operators as shown in Eq.~(\ref{eq:operators-assignment}). Secondly, when the
operators assigned as outgoing from some class $c$ correspond to two independent
subspace decompositions all operators are additionally multiplied by the factor
$1/\sqrt{2}$. In this case classes 0 and 2 utilize both $x$ and $z$
decomposition operators and thus the factor is present in $A_k^{(0,j)}$ and
$A_k^{(2,j)}$ operators.

This gives us the following operator assignment, where $A^{(i,j)}$ is the
operator assigned to every transition from class $i$ to $j$:
\begin{alignat}{4}
A^{(0, 1)}_1 &= B_z/2, & \quad
A^{(1, 0)}_1 &= C_z/\sqrt{6}, & \quad
A^{(1, 2)}_1 &= C_z/\sqrt{6}, & \quad
A^{(2, 1)}_1 &= B_x/2 \label{eq:operators-assignment}, \\ \nonumber
A^{(0, 2)}_1 &= C_x/2, & \quad
A^{(2, 0)}_1 &= C_z/2, & \quad
A^{(1, 3)}_1 &= B_z/\sqrt{6}, & \quad
A^{(3, 1)}_1 &= B_x, \\ \nonumber
A^{(2, 3)}_1 &= B_x/\sqrt{8}, & \quad
A^{(3, 2)}_1 &= C_x/\sqrt{2}, & \quad
A^{(0, 3)}_1 &= B_z/\sqrt{8}, & \quad
A^{(3, 0)}_1 &= C_x/\sqrt{2}, \\ \nonumber
A^{(2, 4)}_1 &= B_z/\sqrt{2}, & \quad
A^{(2, 4)}_2 &= C_x/\sqrt{2}, & \quad
A^{(4, 2)}_1 &= C_x, & \quad
& \\ \nonumber
A^{(0, 4)}_1 &= C_z/2, & \quad
A^{(0, 4)}_2 &= B_x/\sqrt{8}, & \quad
A^{(4, 0)}_1 &= B_x/\sqrt{2}, & \quad
& \\ \nonumber
A^{(0, 0)}_1 &= B_x/\sqrt{8}. & \quad
&&
&&
& \nonumber
\end{alignat}
The class numbers correspond to those shown in Fig.~\ref{fig:network-classes}. 
The initial state is $\rho_0 = \frac14 \1_{\mathbb{C}^4}$.

\begin{figure}[ht!]
\centering
\includegraphics[width=0.8\textwidth]{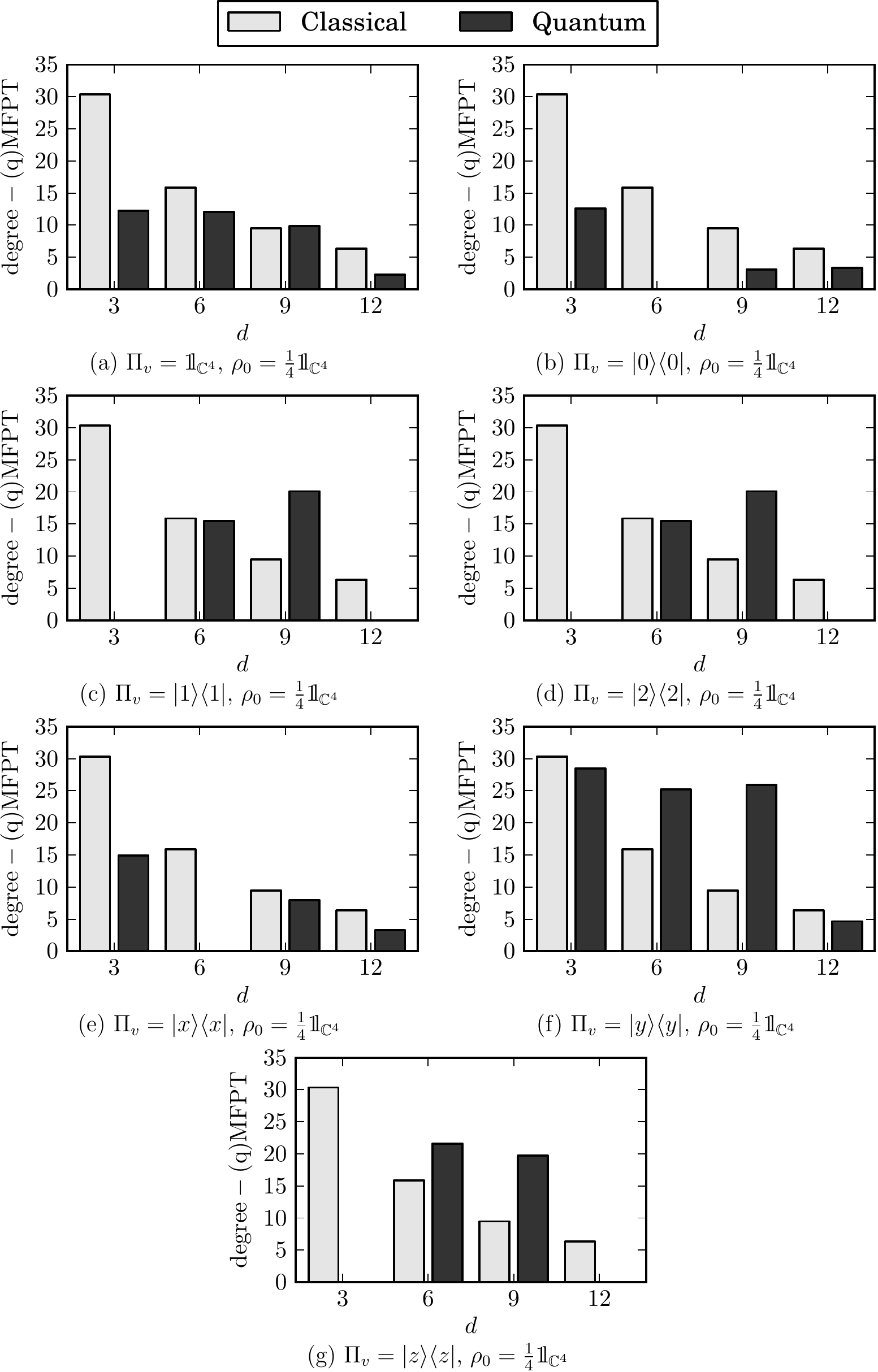}
\caption{{\bf Degree-(q)MFPTs conditioned on the view operator.} A missing
bar indicates that the appropriate vertices are unreachable under the given
view operator. Panel (a) $\Pi_v = \1$, panel (b) $\Pi_v =
\ket{0}\bra{0}$, panel (c) $\Pi_v = \ket{1}\bra{1}$, panel (d) $\Pi_v =
\ket{2}\bra{2}$, panel (e) $\Pi_v = \ket{x}\bra{x}$, panel (f) $\Pi_v =
\ket{y}\bra{y}$, panel (g) $\Pi_v =
\ket{z}\bra{z}$.}\label{fig:MFPT-generations-2}
\end{figure}

\begin{figure}[ht!]
\centering
\includegraphics[width=0.8\textwidth]{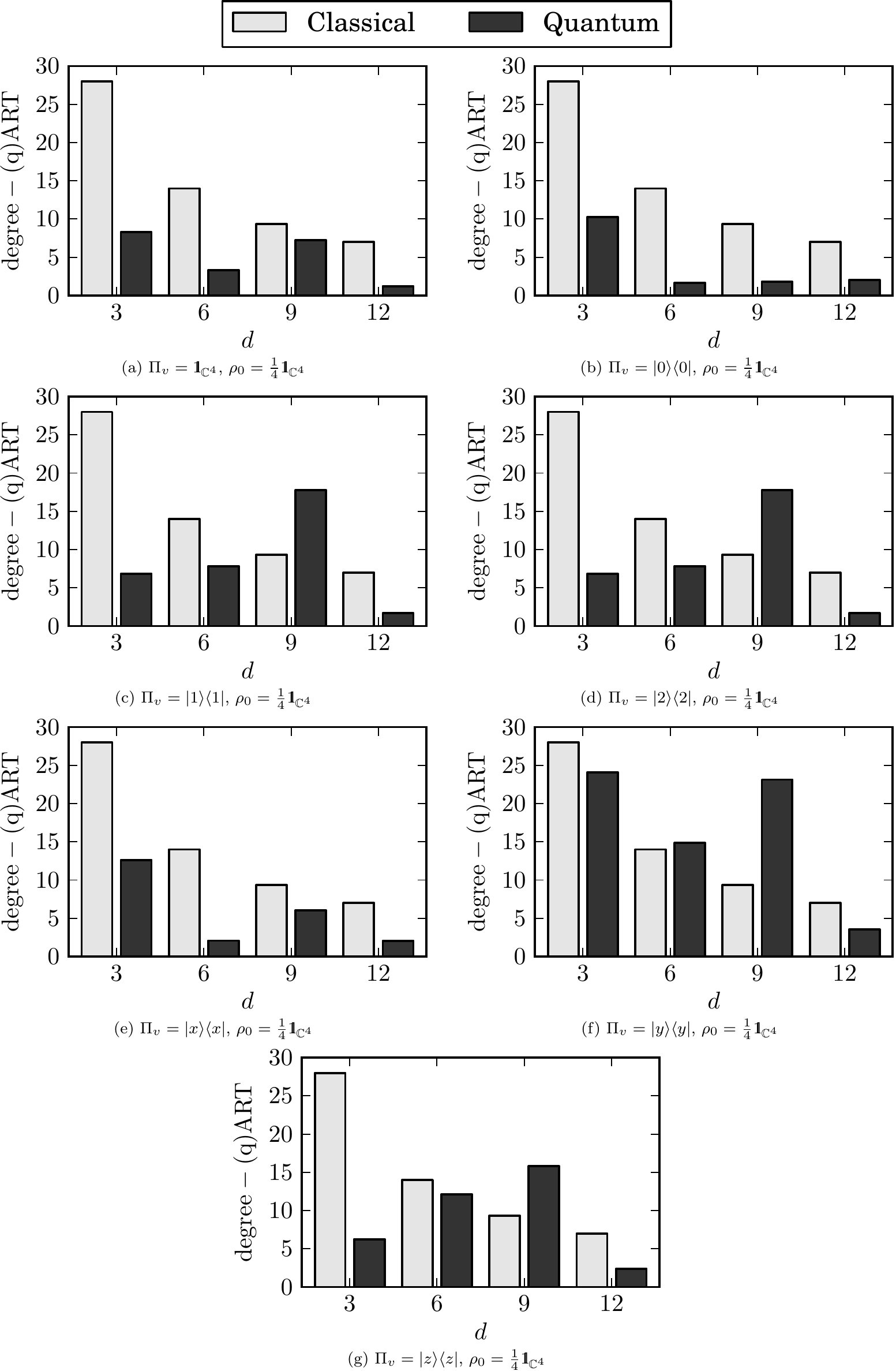}
\caption{{\bf Degree-(q)ARTs conditioned on the view operator.} Panel (a)
$\Pi_v = \1$, panel(b) $\Pi_v = \ket{0}\bra{0}$, panel(c) $\Pi_v =
\ket{1}\bra{1}$, panel(d) $\Pi_v = \ket{2}\bra{2}$, panel(e) $\Pi_v =
\ket{x}\bra{x}$, panel(f) $\Pi_v = \ket{y}\bra{y}$, panel(g) $\Pi_v =
\ket{z}\bra{z}$.}\label{fig:ART-generations-2}
\end{figure}

The numerical results are shown in Figs~\ref{fig:MFPT-generations-2} and
\ref{fig:ART-generations-2}. This time we obtain qMFPTs conditioned on the view
operator which are significantly different from the classical ones. Furthermore,
the ARTs are no longer monotonic functions of the vertex degree $d$ and the
non-classicality is present regardless of the view operator. Moreover, some
positions become unreachable when the view operator is applied.

\section*{Conclusions}\label{sec:final}
The main contribution of this work is the introduction of a generalized model of
open quantum walks, that is derived from the idea of Quantum Markov Chains. We
apply this model to study the evolution of quantum walks on Apollonian networks
that provides some insight on the role of the network properties on the
resulting quantum dynamics. We have also provided definitions of mean first
passage time and average return time for generalized open quantum walks. We have
calculated these quantities for several examples and compared them with the
classical case.

We have shown illustrative set-ups of exciton transport in Apollonian networks
which can lead to very non-trivial behaviour compared to ordinary quantum walks.
In some cases we are able to recover the classical behaviour, although in
general the model allows for much richer walker behaviour. Hence, the open
quantum walk model can be used to explain non-trivial behaviour not only in
linear, but also in more complex topologies of the underlying graphs.
Furthermore, we have studied mean first passage times and average return times
in this set-up. These results differ significantly from a classical walk on
these networks. The results allow us theeasy creation of walks that visit
certain vertices after a given time or omit a selected subset of vertices.

\section*{Acknowledgments}
Work by {\L}P was supported by the Polish Ministry of Science and Higher
Education under the project number IP2012 051272. PS was supported by the Polish
Ministry of Science and Higher Education within ``Diamond Grant'' Programme
under the project number 0064/DIA/2013/42. JAM would like to acknowledge the
support by the Polish National Science Centre under the research project
UMO-2011/03/D/ST6/00413. Research of PG was supported by the Grant N N516 481840
financed by Polish National Science Centre. The authors would like to thank an
anonymous reviewer for his comments.

\bibliography{apollonian_open_walks}

%
%
%

\section*{Figure Legends}
%

\clearpage

\section*{Tables}
%
%
%

\begin{table}[ht!]
\centering
\includegraphics{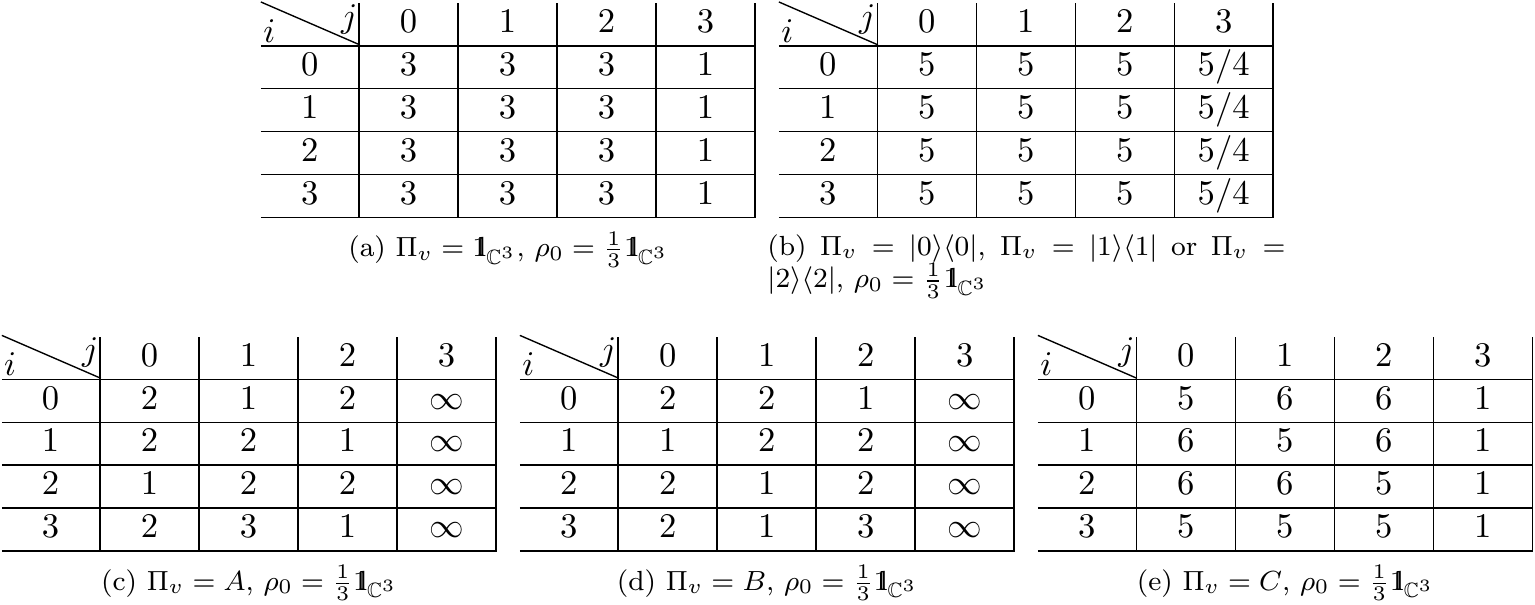}

\caption{qMFPTs from vertex $i$ to vertex $j$ (off-diagonal elements) and qARTs
(diagonal elements) for an open quantum walk on a network shown in
Figure~\ref{fig:walk-4-nodes} conditioned on measurements: panel (a)
$\Pi_\mathrm{v}=\1_{\mathbb{C}^3}$, panel (b) $\Pi_\mathrm{v}=\ketbra{0}{0}$,
$\ketbra{1}{1}$ or $\ketbra{2}{2}$, panel (c) $\Pi_\mathrm{v}=A$, panel (d)
$\Pi_\mathrm{v}=B$, panel (e) $\Pi_\mathrm{v}=C$.}\label{tab:mfpt-4-nodes}
\end{table}

\begin{table}[ht!]
\centering
\includegraphics{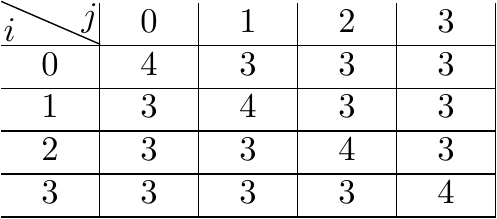}
\caption{MFPTs from vertex $i$ to vertex $j$ (off-diagonal elements) and ARTs
(diagonal elements) for classical random walk on a network shown in
Figure~\ref{fig:walk-4-nodes}.}\label{tab:mfpt-4-nodes-classical}
\end{table}

\begin{table}[ht!]
\centering
\includegraphics{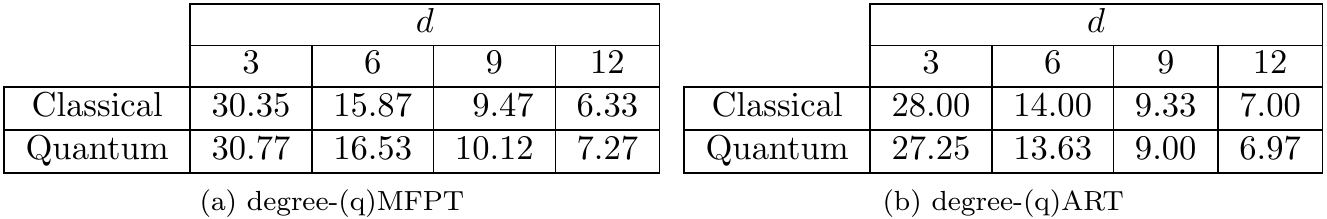}
\caption{{\bf Degree-(q)MFPTs and degree-(q)ARTs for the classical and quantum
for a walk on the Apollonian network of the third generation.} The TOM
assignment is described in the text. Here, we put $\Pi_v = \1_{\mathbb{C}^3}$,
$\rho_0=\ketbra{x}{x}$. We obtain the behaviour different from the classical
case. Notice that the degree-qART does not scale as
$\frac{1}{d}$.}\label{tab:almost-classical}
\end{table}
\clearpage

\section*{Supporting Information Legends}
%
%

\end{document}